\documentclass{article}
\usepackage{arxiv}

\usepackage[utf8]{inputenc}
\usepackage[T1]{fontenc}
\usepackage{physics}
\usepackage{siunitx}
\usepackage{hyperref}       % hyperlinks
\usepackage{url}            % simple URL typesetting
\usepackage{booktabs}       % professional-quality tables
\usepackage{amsfonts}       % blackboard math symbols
\usepackage{nicefrac}       % compact symbols for 1/2, etc.
\usepackage{microtype}      % microtypography
\usepackage{graphicx}
\usepackage{doi}
\usepackage[version=3]{mhchem} % Formula subscripts using \ce{}
\usepackage{subcaption}

\newcommand*{\figref}[2][]{%
  \hyperref[{#2}]{%
    \ref*{#2}%
    \ifx\\#1\\%
    \else
      #1%
    \fi
  }%
}

\title{Neural network enabled wide field-of-view imaging with hyperbolic metalenses}

\date{} 					% Or removing it

\author{%
\href{https://orcid.org/0000-0001-5160-7628}{\includegraphics[scale=0.06]{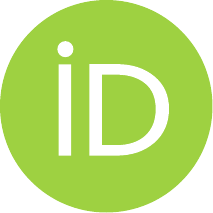}\hspace{1mm}\textbf{Joel Yeo}$^{1,2,3}$}
\quad
\href{https://orcid.org/0000-0002-5733-3952}{\includegraphics[scale=0.06]{orcid.pdf}\hspace{1mm}\textbf{Deepak K. Sharma}$^{3}$}
\quad
\href{https://orcid.org/0000-0001-6420-1440}{\includegraphics[scale=0.06]{orcid.pdf}\hspace{1mm}\textbf{Saurabh Srivastava}$^{3}$}
\quad
\href{https://orcid.org/0000-0003-4609-173X}{\includegraphics[scale=0.06]{orcid.pdf}\hspace{1mm}\textbf{Aihong Huang}$^{3}$}
\\
\href{https://orcid.org/0000-0002-0098-5159}{\includegraphics[scale=0.06]{orcid.pdf}\hspace{1mm}\textbf{Emmanuel Lassalle}$^{3}$}
\quad
\href{https://orcid.org/0000-0002-0848-552X}{\includegraphics[scale=0.06]{orcid.pdf}\hspace{1mm}\textbf{Egor Khaidarov}$^{3}$}
\quad
\textbf{Keng Heng Lai}$^{4}$
\quad
\href{https://orcid.org/0000-0002-7691-0196}{\includegraphics[scale=0.06]{orcid.pdf}\hspace{1mm}\textbf{Yuan Hsing Fu}$^{4}$}
\quad
\href{https://orcid.org/0000-0002-8886-510X}{\includegraphics[scale=0.06]{orcid.pdf}\hspace{1mm}\textbf{N. Duane Loh}$^{1,2,5}$} 
\\
\href{https://orcid.org/0000-0002-7622-8939}{\includegraphics[scale=0.06]{orcid.pdf}\hspace{1mm}\textbf{Arseniy I. Kuznetsov}$^{3,*}$}
\quad
\href{https://orcid.org/0000-0001-7836-681X}{\includegraphics[scale=0.06]{orcid.pdf}\hspace{1mm}\textbf{Ramon Paniagua-Dominguez}$^{3,+}$} 
\\
$^1$NUS Graduate School for Integrative Sciences and Engineering Programme, National University of\\Singapore, 119077 Singapore, Singapore, \\ 
$^2$Department of Physics, National University of Singapore, 117551 Singapore, Singapore,\\
$^3$Institute of Materials Research and Engineering (IMRE), Agency for Science, Technology and
Research\\(A*STAR), 2 Fusionopolis Way, Innovis \#08-03, 138634 Singapore, Singapore,\\
$^4$Institute of Microelectronics (IME), Agency for Science, Technology and
Research\\(A*STAR), 2 Fusionopolis Way, Innovis \#08-02, 138634 Singapore, Singapore,\\
$^5$Department of Biological Sciences, National University of Singapore, 117558 Singapore, Singapore,\\
\\
$^+$\texttt{Corresponding author: Ramon\_Paniagua@imre.a-star.edu.sg}
\\
$^*$\texttt{Corresponding author: Arseniy\_Kuznetsov@imre.a-star.edu.sg}
}

\hypersetup{
pdftitle={Neural network enabled wide field-of-view imaging with hyperbolic metalenses},
pdfsubject={},
pdfauthor={Joel Yeo, Deepak, K. Sharma, Saurabh Srivastava, Aihong Huang, Emmanuel Lassalle, Egor Khaidarov, Keng Heng Lai, Yuan Hsing Fu, N. Duane Loh, Ramon Paniagua-Dominguez, and Arseniy I. Kuznetsov},
}

\begin{document}

\maketitle

\begin{abstract}
The ultrathin form factor of metalenses makes them highly appealing for novel sensing and imaging applications.
Amongst the various phase profiles, the hyperbolic metalens stands out for being free from spherical aberrations and having one of the highest focusing efficiencies to date.
For imaging, however, hyperbolic metalenses present significant off-axis aberrations, severely restricting the achievable field-of-view (FOV).
Extending the FOV of hyperbolic metalenses is thus feasible only if these aberrations can be corrected.
Here, we demonstrate that a Restormer neural network can be used to correct these severe off-axis aberrations, enabling wide FOV imaging with a hyperbolic metalens camera.
Importantly, we demonstrate the feasibility of training the Restormer network purely on simulated datasets of spatially-varying blurred images generated by the eigen-point-spread function (eigenPSF) method, eliminating the need for time-intensive experimental data collection.
This reference-free training ensures that Restormer learns solely to correct optical aberrations, resulting in reconstructions that are faithful to the original scene.
Using this method, we show that a hyperbolic metalens camera can be used to obtain high-quality imaging over a wide FOV of $54^\circ$ in experimentally captured scenes under diverse lighting conditions.
\end{abstract}

\section{Introduction}
\label{sec:introduction}

Metasurfaces have emerged as a transformative technology in optics due to their potential to replace, or even outperform, traditional optical components with ultra-thin, multi-functional ones. 
Within the field, the metasurface counterparts of traditional lenses (so-called metalenses) are particularly attractive as these are the most ubiquitous elements in optical systems, usually taking the vast majority of space and weight. 
Unlike traditional bulky lenses, metalenses utilize nanoscale structures to manipulate the fundamental properties of light (typically the phase) locally and abruptly, making them invaluable for applications in imaging, sensing, and optical metrology \cite{Kuznetsov2024-np, Ha2024}. 
The capability of metalenses to replicate complex phase profiles while remaining ultrathin also offers significant advantages over their bulky counterparts, where freeform optics is usually expensive and difficult to manufacture \cite{Kuznetsov2016-gy, Genevet2017-cd, Pan2022-uk}. 

Within the different metalens designs explored by the community, the one that imparts a hyperbolic phase profile in the incident beam is particularly attractive as it is free from spherical (and any other spatial) aberrations when illuminated on-axis \cite{Pan2022-uk}.
In addition, the focusing efficiency of these metalenses remains the highest demonstrated to date, making them commonly used for light-focusing applications, including those requiring high-numerical apertures (NA) \cite{aieta2012aberration, Paniagua-Dominguez2017-we, Khorasaninejad2016-vx, Fan2018-qs, Liang2018-yn, Huang2019-fd}.
However, while the hyperbolic lens is theoretically diffraction-limited along the optical axis, it presents strong off-axis aberrations, translating into a point-spread function (PSF) that rapidly deteriorates as the angle of incident light departs from normal \cite{Liang2019-de}.
In an imaging experiment, this causes the resultant image to be aberrated with a spatially varying blur, which traditional deblurring methods such as the Wiener filter \cite{Wiener2019-je} cannot remove. 
As a consequence, these off-axis aberrations severely limit the usable field-of-view (FOV) of hyperbolic metalenses and, therefore, their use in imaging applications. 

To circumvent this issue and expand the FOV of metalenses, the community has explored alternative phase profiles, such as the quadratic one \cite{pu2017nanoapertures, martins2020metalenses, Sharma2024-wz, Baranikov2023-hz, Lassalle2021-wl, Luo2022-gs, yang2023wide}, or multi-element configurations (doublets, triplets, or other lens arrays)\cite{martins2022fundamental, arbabi2016miniature, wirth2025wide, shalaginov2020single, Groever2017-rk}.
While these are indeed able to provide a wide FOV (up to even $180^\circ$ in some cases), they come at the cost of spherical aberrations and poor efficiencies (in the case of quadratic phase profiles) or fabrication complexity and overall system size (in the case of doublets or systems with an aperture). 

As a result, part of the community is now turning their attention to the possibility of correcting this issue on the software side rather than the hardware one. 
In this regard, iterative deconvolution algorithms have been recently introduced to correct for such spatially varying aberrations \cite{Yeo2024-pm}. 
These, however, are typically slow and prone to reconstruction artifacts \cite{Baranikov2023-hz}.
These algorithms are also sensitive to noise and require precise calibration of the spatially-varying PSFs which is challenging in practical applications.
Over recent years, deep-learning algorithms have been increasingly applied to remove aberrations from metalens images \cite{Dun2020-uh, Tseng2021-ei, Tan2021-dk, Fan2022-dg, Hu2023-ys, Zhang2024-gu, Pinilla2023-ll, Maman2023-ci, Liu2024-fe, Cheng2024-ie}.
Their fast inference speed, combined with robustness against noise and experimental errors, make them highly appealing and successful for metalens imaging postprocessing.
However, many demonstrations of deep-learning deblurring are reference-based, requiring tedious curation of experimental datasets of measurement and ground truth pairs.
This could also result in overfitting to specific imaging conditions, such as lighting, magnification, alignment, and other experimental parameters for which the experimental dataset was collected.
As such, these trained networks would not be readily extended to deblur images under different imaging conditions.

Here, we present a neural network-enabled, reference-free hyperbolic metalens camera for wide FOV imaging.
In particular, we employ a Restormer neural network to correct the severe off-axis aberrations of these type of lenses, ultimately enabling aberration-free imaging over $54^\circ$.
Reference-free training of the network is performed by simulating datasets of spatially-varying blurred images using the eigenPSF method \cite{Yeo2024-pm}.
This eliminates the need for time-consuming curation of experimental datasets and also ensures that the trained network only removes optical aberrations without overfitting to specific imaging conditions.
We demonstrate that our hyperbolic metalens camera delivers robust imaging performance in low-light conditions, during close-up photography, and under diverse lighting directions and occlusions.

\begin{figure*}[t]
    \centering
    \includegraphics[width=1\textwidth]{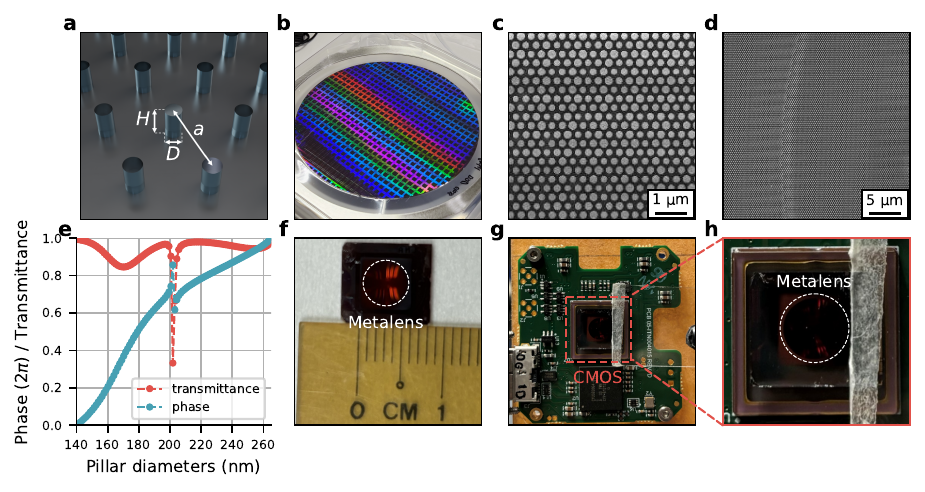}
    \caption{Fabrication of the hyperbolic metalens and the imaging setup.
    (a) Schematic of the hexagonal unit cell of the metalens with a lattice constant of $a=$ \SI{350}{\nm}, consisting of cylindrical nanopillars with a height of $H=$ \SI{500}{\nm} and diameters, $D$, ranging from \SI{140}{\nm} to \SI{264}{\nm}.
    (b) Optical image of a wafer with an array of metalenses patterned using deep UV immersion photolithography.
    (c,d) Scanning electron micrographs of the metalens depicting the patterned a-Si nanopillars on a glass substrate.
    (e) The simulated phase and transmittance of uniform a-Si nanopillar arrays with height of 500nm and diameters ranging from 140 to \SI{264}{\nm} with a step size of \SI{2}{\nm}.
    (f) Optical image of the fabricated \SI{5}{\mm} diameter metalens.
    (g) Optical image of the hyperbolic metalens camera used in imaging, where the (h) metalens (white circle) is mounted directly in front of the CMOS detector (red square).
    }
    \label{fig:lens}
\end{figure*}

\section{Results}
\label{sec:results}

\subsection{Design, fabrication and optical characterization of the metalens}

The hyperbolic metalens used in this work has a phase profile given by the expression
\begin{align}
    \phi\qty(r) = \frac{2\pi}{\lambda}\qty(f - \sqrt{r^2 + f^2}),
\end{align}
where $r$ is the radial distance from the center of the lens, and $\lambda$ and $f$ are the design wavelength and focal length, respectively.
The fabricated metalens has a diameter of $D = $ \SI{5}{\mm} and $f = $ \SI{1.813}{\mm}, designed at a working wavelength of $\lambda = $ \SI{850}{\nm} with a numerical aperture of $\text{NA}=0.81$.
The (wrapped) hyperbolic phase profile was mapped using amorphous silicon (a-Si) nanopillars with a circular cross-section (to maintain polarization-insensitive response) on a fused silica substrate.
These pillars are arranged in a hexagonal lattice (lattice constant of \SI{350}{\nm}) and have a fixed height of \SI{500}{\nm} and diameters in the range of 140 - \SI{264}{\nm} (Fig.~\figref[a,b,c,d]{fig:lens}). 
The simulated transmittance and phase as a function of the pillar diameter are also plotted in Fig.~\figref[e]{fig:lens}.

\begin{figure*}[!t]
    \centering
    \includegraphics[width=0.8\textwidth]{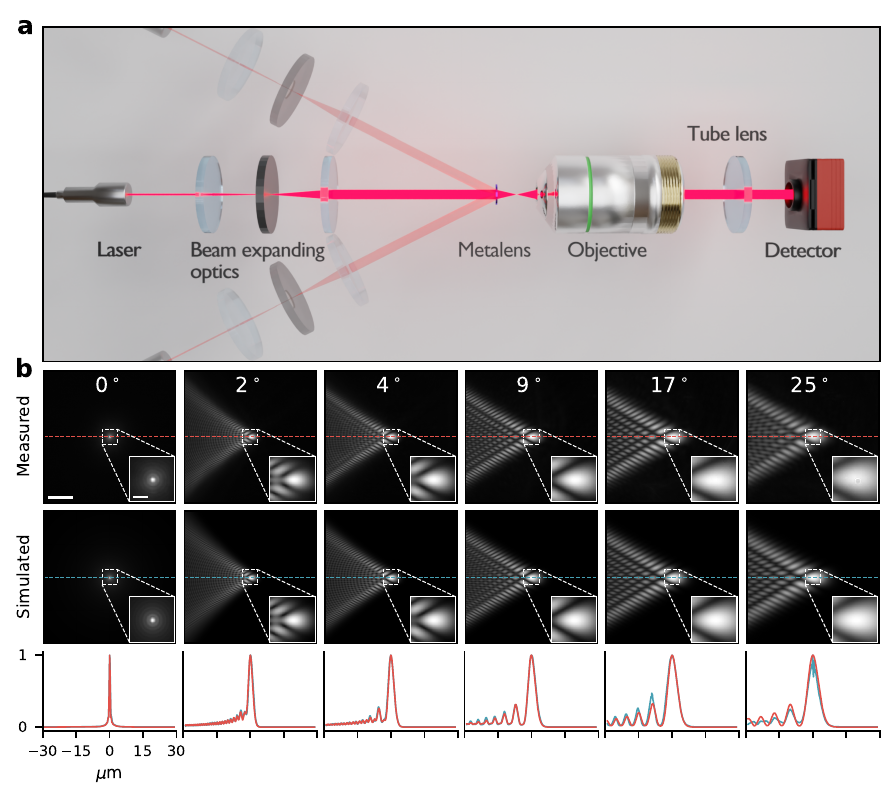}
    \caption{Characterization of the hyperbolic metalens.
    (a) Optical setup for angles of incidence dependent optical characterization of the hyperbolic metalens PSFs.
    The laser (\SI{850}{\nm} wavelength) output is expanded using two lenses and an aperture to distribute the intensity uniformly across the metalens.
    These are mounted on a rotating arm of a goniometer that enables different AOI illumination on the metalens.
    The metalens focuses the collimated laser at the focal plane, which is then imaged onto the CMOS detector using an objective lens (Olympus, MPLAPON100X) and a tube lens (\SI{150}{\mm} focal distance).
    (b) (top) Experimentally measured PSFs at different angles of incidence compared to (middle) simulated PSFs for the hyperbolic metalens.
    We use a log-normalized colormap for better visualization of these PSFs.
    The scalebars (white) have sizes of \SI{10}{\um} and \SI{2}{\um} for the main image and its inset, respectively.
    (bottom) The horizontal line profiles of the measured (red) and simulated (blue) PSFs.
    }
    \label{fig:characterization}
\end{figure*}

The samples are fabricated using a 12-inch, deep ultraviolet (UV) immersion photolithography scanner (see details in Methods) and optically characterized using a goniometric optical setup (Fig.~\figref[a]{fig:characterization}).
This characterization setup, which has a calculated magnification of 83.3 (resulting in an effective detector pixel size of \SI{41.4}{\nm}) allows imaging the PSFs of the fabricated hyperbolic metalens at various angles of incidence (AOI).
Fig.~\figref[b]{fig:characterization} compares these measurements against theoretical PSFs calculated with Fourier optics simulations (details in Supplementary Information).
As can be seen, there is a close match between the measured and simulated PSFs, with minor discrepancies likely attributed to fabrication errors.

\subsection{Hyperbolic metalens camera}
The hyperbolic metalens camera comprises only two components (Fig.~\figref[g,h]{fig:lens}): the metalens and a complementary metal oxide semiconductor (CMOS) detector (Thorlabs, Zelux-CS165MU), resulting in an ultra-compact design.
The scene is illuminated with a light-emitting diode (LED) with a dominant wavelength of \SI{850}{\nm} (Thorlabs M850L3) and bandwidth of \SI{30}{\nm} (setup figure in Supplementary Information).
The hyperbolic metalens, mounted at a distance $f =$ \SI{1.813}{\mm} from the detector, focuses the illuminated scene onto the sensor to form the image.
In this work, we used only the detector's central $512 \times 512$ pixels, corresponding to an angular field-of-view (FOV) of $54^\circ$, due to memory constraints during network training.

\subsection{Restormer deblurring}

\begin{figure*}[!t]
    \centering
    \includegraphics[width=0.8\textwidth]{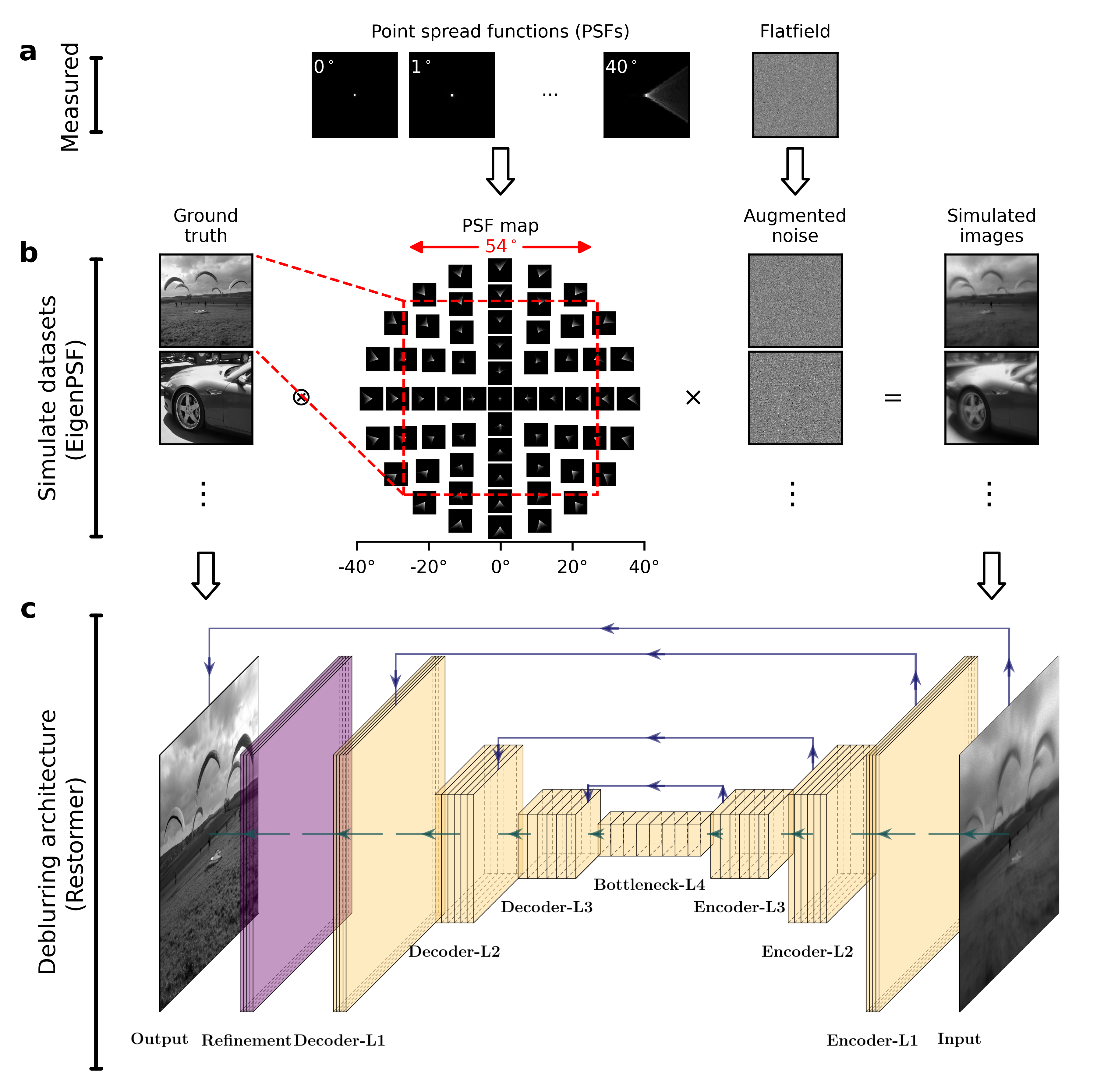}
    \caption{Schematic for computational deblurring using the Restormer architecture trained on eigenPSF-simulated images. 
    (a) The hyperbolic metalens is characterized using the measured PSFs, and the flatfield records the noise profile of the imaging system. 
    (b) Simulated images are obtained using the eigenPSF method to apply spatially varying blur on the ground truth dataset, and further corrupted by noise created by augmenting the flatfield measurement.
    (c) The Restormer architecture is used to deblur and denoise the images. Illustration created using PlotNeuralNet \cite{haris_iqbal_2018_2526396}.}
    \label{fig:schematic}
\end{figure*}

Fig.~\ref{fig:schematic} shows the schematic of our computational deblurring approach for hyperbolic metalens imaging.
Using our imaging setup (see Supplementary Information), we first measured the PSFs at different AOIs ranging from $0^\circ$ to $40^\circ$ (Fig.~\figref[a]{fig:schematic}).
Note that these imaging PSFs are different from the measured PSFs depicted in Fig.~\figref[b]{fig:characterization} as there is no external magnification in the imaging setup.
These imaging PSFs are computationally rotated to populate a PSF map that covers the full extent of an image corresponding to an angular FOV of $54^\circ$ as shown in Fig.~\figref[b]{fig:schematic}.
The eigenPSF method \cite{Yeo2024-pm} (see Supplementary Information) uses this PSF map to simulate spatially-varying blur applied to ground truth images from Google's Open Images dataset \cite{OpenImages, OpenImages2}.
Each simulated blurred image is corrupted with noise by augmenting a measured flatfield through random rotations and flips.
The total time taken to simulate 3500 noisy and blurred images on a single NVIDIA L40 GPU was approximately 10 minutes.

A Restormer network \cite{Zamir_2022_CVPR} is trained using these 3500 simulated images as input, and their corresponding ground truth images as the desired output (Fig.~\figref[c]{fig:schematic}).
We use the default parameters and loss functions described in the original paper \cite{Zamir_2022_CVPR} for the Restormer network, except reducing the number of channels to $[36, 72, 144, 288]$ for layers L1 to L4 respectively due to GPU memory constraints.
This constitutes a total of 14.8 million trainable parameters in our Restormer network.
Using 4 NVIDIA L40 GPUs, training for 200 epochs with a batch size of 1 took approximately 48 hours.

\begin{figure*}[h]
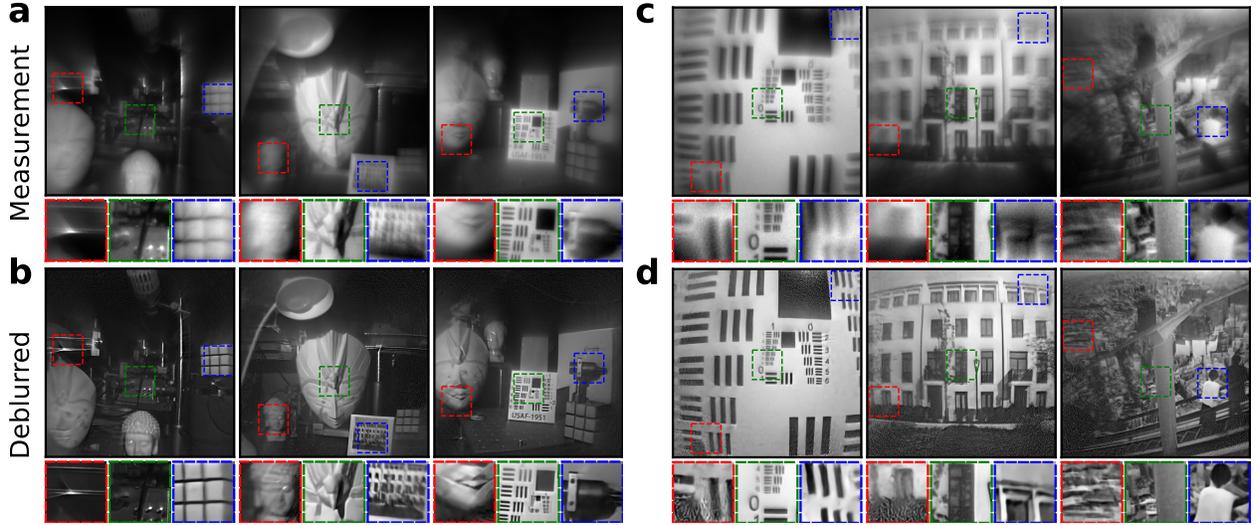

     \centering
     \begin{subfigure}[b]{0.495\textwidth}
         \centering
         \includegraphics[width=\textwidth]{images/deblurred_3scenes.png}
     \end{subfigure}
     \hfill
     \begin{subfigure}[b]{0.495\textwidth}
         \centering
         \includegraphics[width=\textwidth]{images/deblurred_3cards.png}
     \end{subfigure}
     \caption{Deblurring images from the hyperbolic metalens camera using the trained, reference-free Restormer network.
     The (a) measured and (b) deblurred images of scenes taken around a lab.
     The (c) measured and (d) deblurred images of printed photos placed before the camera. 
     All images have the same angular FOV of $54^\circ$.}
     \label{fig:deblur}
\end{figure*}

\begin{figure*}[!h]
     \centering
     \begin{subfigure}[b]{0.495\textwidth}
         \centering
         \includegraphics[width=\textwidth]{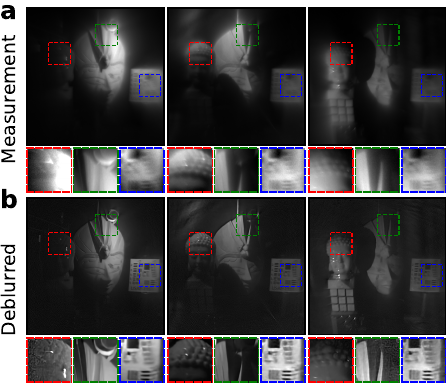}
     \end{subfigure}
     \hfill
     \begin{subfigure}[b]{0.495\textwidth}
         \centering
         \includegraphics[width=\textwidth]{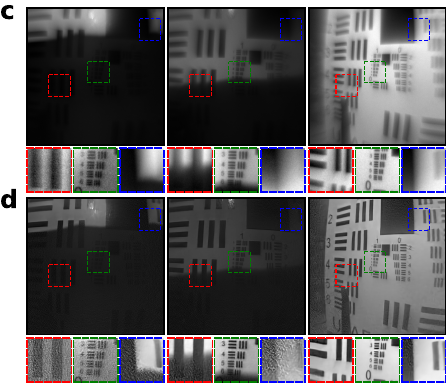}
     \end{subfigure}
     \caption{Deblurring images from the hyperbolic metalens camera with varying illumination direction and obstructions.
     The (a) measured and (b) deblurred images of the same lab scene under different lighting.
     The (c) measured and (d) deblurred images of a printed USAF card, where the card was tilted in the last column. 
     All images have the same angular FOV of $54^\circ$.}
     \label{fig:deblur_varillum}
\end{figure*}

Fig.~\ref{fig:deblur} shows the raw measurements from our hyperbolic metalens camera and the corresponding results of Restormer deblurring on the images (see Supplementary Information for more results).
The characteristic aberrations due to the hyperbolic lens phase profile are evident in the sharp features at the center of the image and the increasing coma at larger incidence angles.
The images here are not diffraction-limited due to the broadband LED used to illuminate the scenes and photos.
The physical size of the detector pixels also limits the resolution of the measurements.

Despite the spatially-varying aberrations in the measurements, the trained Restormer network is able to deblur the full FOV of the images in real-time ($\sim$\SI{50}{\ms} per image), recovering features even toward the edges of the images.
By using a reference-free dataset, we avoid overfitting to specific imaging conditions during the training of the Restormer network.
% This is evident in the similar quality of deblurring from both measurements of lab scenes (dimmer) and printed photos (brighter) in Fig.~\ref{fig:deblur}.
This is further demonstrated in Fig.~\ref{fig:deblur_varillum} where under varying illumination directions and obstructions, our trained Restormer is still able to recover features even with low lighting at various regions of both the scene and printed USAF card.

\begin{figure*}[h]
    \centering
    \includegraphics[width=0.8\linewidth]{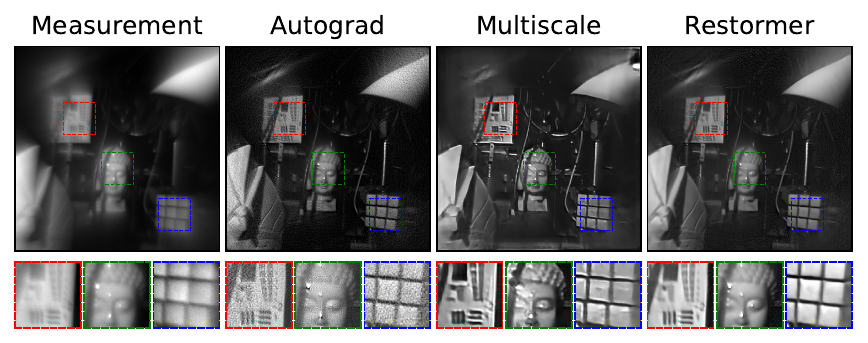}
    \caption{Comparing deblurring performance using various algorithms. The Restormer network surpasses both Autograd and Multiscale methods in both spatially-varying deblurring capabilities as well as suppressing noise.}
    \label{fig:compare}
\end{figure*}

Fig.~\ref{fig:compare} further demonstrates the improved quality of deblurring from our trained Restormer network over other existing state-of-the-art approaches.
The first of which is an Autograd implementation of the eigenCWD algorithm \cite{Yeo2024-pm} which utilizes PyTorch's \cite{Paszke2019-aa} inbuilt automatic differentiation engine to perform optimization instead of using analytical gradients (details in Supplementary Information).
The reconstruction from this iterative approach in Fig.~\ref{fig:compare} is contaminated with noisy artifacts as it only accounts for spatially-varying blur and not the noise characteristics of the sensor.
In addition, as an iterative algorithm, the Autograd implementation of eigenCWD is incapable of real-time deblurring ($\sim$\SI{1}{\minute} per image on a single NVIDIA L40 GPU).
Using the same dataset, we also trained a Multiscale neural network architecture \cite{Nah2017-ot} which has recently been used in image reconstruction applications for metalenses \cite{Hu2023-ys, Zhang2024-gu} (details in Supplementary Information).
However, we observe residual smeared artifacts in the output of the Multiscale network, likely attributed to the presence of noise in the training dataset which the network is unable to fully remove.
This suggests that the Restormer network remains robust against noise and demonstrates improved performance in spatially-varying deconvolution over existing state-of-the-art methods.

\section{Conclusion}
\label{sec:conclusion}
In this work, we have demonstrated wide FOV imaging with a hyperbolic metalens camera.
By using the eigenPSF method as an efficient forward model to simulate the metalens' spatially-varying blur computationally, we circumvent the need for experimental curation of datasets for training a deblurring neural network.
In addition, the Restormer network used for postprocessing the images enables real-time aberration correction (after training) compared to time-consuming iterative algorithms, and additionally remains robust against noise and experimental errors.

Our findings suggest that the field of view (FOV) in hyperbolic metalens imaging could be further extended by leveraging advances in computational power to train on larger image sizes.
Additionally, the diffraction-limited resolution of the hyperbolic lens along the optical axis remains underutilized due to current limitations in detector pixel sizes and the large bandwidth of the illumination source. 
With future improvements in hardware, this work has the potential to open new pathways toward achieving high-resolution, wide-FOV imaging with hyperbolic metalenses.

\section{Methods}
\label{sec:methods}
\subsection{Fabrication of Metalens}

A \SI{193}{\nm} argon fluoride (ArF) deep-ultraviolet (DUV) immersion photolithography process combined with a dry etching process is used to fabricate the hyperbolic metalenses.
The metalens comprises millions of amorphous silicon (a-Si) nanopillars, which are patterned on a \SI{350}{\nm}-thick a-Si film deposited using plasma-enhanced chemical vapor deposition (PECVD) on a 12-inch fused silica wafer.
The wafer was then diced into small coupons, and the individual a-Si metalenses were subsequently etched using a dry etching process.
Hence, the metalens pattern was transferred from the photoresist to the a-Si film, forming a-Si nanopillars.
The \SI{480}{\nm} a-Si pillars were etched in multiple smaller steps instead of a continuous step.
This process is recommended, especially when performing deep etching with high aspect ratio pillars.
After every etching step, the chamber undergoes a 5-minute cooling period before the next step.
This not only provides smoother sidewalls but also protects the pillars from undercutting.
A residue layer of \ce{SiO2} (\SI{30}{\nm}) remains on top after a-Si etching as a part of the etching hard mask, but it possesses no hindrance to the optical performance, and therefore, we do not remove it.

\paragraph{Funding}    
This work was supported by the A*STAR Graduate Scholarship, the AME Programmatic Grant, Singapore, under Grant A18A7b0058., and the Early Career Research Award from the National University of Singapore (NUS).

\paragraph{Acknowledgements}    
The authors would like to acknowledge the computational resources provided by the NUS Centre for Bio-Imaging Sciences.

\paragraph{Data availability}
The code and dataset used in this work are available at \url{https://doi.org/10.5281/zenodo.14746073}.
    
\paragraph{Supplemental document}
See Supplementary Information for supporting content.

\bibliographystyle{unsrt}
\bibliography{main}

\newpage

\appendix

\begin{center}
    \textbf{{\huge Supplementary Information}}
\end{center}
\numberwithin{equation}{section}
\setcounter{equation}{0}

\section{Fourier optics simulation of PSFs for characterization}
The measured PSFs to characterize the fabricated hyperbolic lens in Fig. 1 have a calculated magnification of 83.3.
The detector has a reported pixel size of \SI{3.45}{\um}, which implies that the measured PSFs have an effective pixel size of $\frac{\SI{3.45}{\um}}{83.3} = 41.4$ nm.
However, because it is computationally expensive to simulate the large lens diameter of $d = \SI{5}{\mm}$ with such a small pixel size, we choose to simulate PSFs which have a larger pixel size of \SI{165.6}{\nm}, and compare them to the 4x binned versions of the measured PSFs, and these are the depicted images in Fig. 2.

The hyperbolic phase profile is defined as
\begin{align}
    \phi\qty(\vb{x}) = \frac{2\pi}{\lambda_0}\qty(f - \sqrt{x^2 + y^2 + f^2})
\end{align}
where $\vb{x} = \qty(x,y)$ are the cartesian coordinates, and $\lambda_0=\SI{850}{\nm}$ and $f=\SI{1.731}{\mm}$ are the designed wavelength and focal length of the metalens.
This phase profile is sampled on a $10001\times10001$ grid with an initial pixel size of \SI{500}{\nm}.
The circular lens function is therefore
\begin{align}
    L\qty(\vb{x}) = 
    \begin{cases}
        \exp[i\phi\qty(\vb{x})], & \text{ for }\sqrt{x^2+y^2} \leq \frac{d}{2}\\
        0, &\text{ otherwise},
    \end{cases}
\end{align}

We simulate a plane wave, $\psi_0$, as
\begin{align}
    \psi_0(\vb{x}) = \exp[ik\qty(x \sin \theta_x + y \sin \theta_y)],
\end{align}
where $k = \frac{2\pi}{\lambda}$ is the wavenumber of the plane wave (can be different from $\lambda_0$), and $\theta_x$ and $\theta_y$ are the angles of incidence with respect to the $x$ and $y$ axes.

This plane wave passes through the metalens, which results in the exitwave, $\psi_\mathrm{exit}$:
\begin{align}
    \psi_\mathrm{exit}\qty(\vb{x}) = \psi_0\qty(\vb{x}) L\qty(\vb{x})
\end{align}
This exitwave is then propagated to the detector plane at a distance $f$ away to form the PSF using the scaled band-limited angular spectrum (BLAS) method \cite{Yu2012-ff}
\begin{align}
    \mathrm{PSF}\qty(\vb{x}) = \abs{\mathcal{P}_f^m\qty{\psi_\mathrm{exit}\qty(\vb{x})}}^2,
\end{align}
where $\mathcal{P}_f^m$ is the BLAS propagator with a magnification of $m \approx 3.019$ such that the final pixel pitch of the simulated PSF is $\frac{\SI{500}{\nm}}{3.019}  \approx  \SI{165.6}{\nm}$.

\begin{figure}[h]
    \centering
    \includegraphics{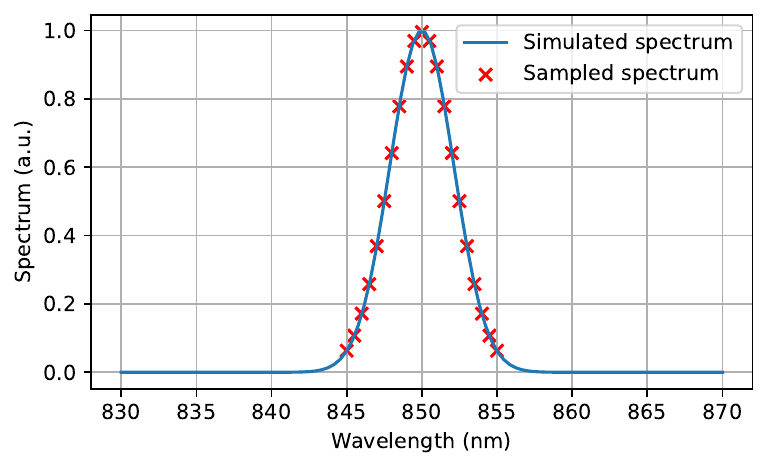}
    \caption{The simulated Gaussian spectrum of the laser with FWHM of \SI{5}{nm}, and the sampled points used to calculate the broadband PSF.}
    \label{fig:spectrum}
\end{figure}
The laser used to characterize the PSFs has a full width-half-maximum (FWHM) of \SI{5}{nm} (Fig.~\ref{fig:spectrum}).
We approximate the laser's spectrum as a Gaussian function centered on \SI{850}{\nm} with a standard deviation of $\frac{\SI{5}{\nm}}{2.355}=\SI{2.123}{\nm}$, where the relationship between the FWHM and standard deviation, $\sigma$, of a normal Gaussian function is
\begin{align}
    \mathrm{FWHM} \approx 2.355 \sigma.
\end{align}
We then sampled 21 equally spaced points from this spectrum between $\qty[\SI{845}{\nm}, \SI{855}{\\nm}]$, and the broadband PSF is simply the incoherent sum of these monochromatic PSFs weighted by their corresponding spectrum value.

\section{Simulating spatially-varying blur with eigenPSF}
Here, we briefly describe the eigenPSF method used to efficiently simulate spatially-varying aberrations.
The full details of this method can be found in \cite{Yeo2024-pm} where open-source code is also available.

For an incoherent imaging system where the point spread function varies as a function of spatial positions, the image formation model is
\begin{align}\label{eq:double}
    g\qty(x,y) = \iint f\qty(u,v)p\qty(u,v,x-u,y-v)\dd{u}\dd{v},
\end{align}
where $g$ is the aberrated image, $\qty(x,y)$ denotes the spatial coordinates of the image plane, $f$ is the object, $\qty(u,v)$ denotes the spatial coordinates of the object plane, and $p$ are the corresponding point-spread functions (PSFs) of the imaging system which vary depending on the point emitter's $\qty(u,v)$ coordinates.
Explicitly computing the double integral in Eq.~\eqref{eq:double} is highly inefficient and slow, and scales like $\order{N^4}$ for an object of size $N\times N$ pixels.

Instead, one can approximate the spatially-varying PSFs as a weighted, linear sum of eigenPSFs
\begin{align}\label{eq:eigenpsf}
    p\qty(u,v,x,y) = \sum_{i=1}^{\infty}a_i\qty(u,v)q_i\qty(x,y),
\end{align}
where $q_i$ are the eigenPSFs and $a_i$ are the eigencoefficients which encode the variation of the PSFs over the extent of the object.
Substituting Eq.~\eqref{eq:eigenpsf} into Eq.~\eqref{eq:double} results in
\begin{align}\label{eq:eigenconv}
    g\qty(x,y) = \sum_{i=1}^\infty\qty[\qty(fa_i)\otimes q_i]\qty(x,y),
\end{align}
where $\otimes$ denotes a convolution operator.
Eq.~\eqref{eq:eigenconv} describes a sum of 2D convolutions, which can be efficiently computed with Fourier transforms based on the convolution theorem.

The eigenPSFs, $q_i$, and eigencoefficients, $a_i$, are numerically calculated by performing an eigendecomposition on a stack of PSFs at various sampled $\qty(u,v)$ locations on the object plane.
The mathematical details can be found in \cite{Yeo2024-pm}.

In this paper, we experimentally measured hyperbolic PSFs at 0$^\circ$, 1$^\circ$, 2$^\circ$, 3$^\circ$, 4$^\circ$, 5$^\circ$, 6$^\circ$, 7$^\circ$, 8$^\circ$, 9$^\circ$, 10$^\circ$, 12$^\circ$, 14$^\circ$, 16$^\circ$, 18$^\circ$, 20$^\circ$, 25$^\circ$, 30$^\circ$, and 40$^\circ$ angles of incidences.
These PSFs were numerically rotated to fully cover the finite extent of the imaged object (see Fig. 3), resulting in a total of 692 PSFs.
We then computed the corresponding eigenPSFs and eigencoefficients from these 692 PSFs and used Eq.~\eqref{eq:eigenconv} to simulate the spatially-varying aberrated images for our training dataset.

\newpage

\section{Modulation transfer function of the hyperbolic metalens}
\begin{figure}[h]
    \centering
    \includegraphics{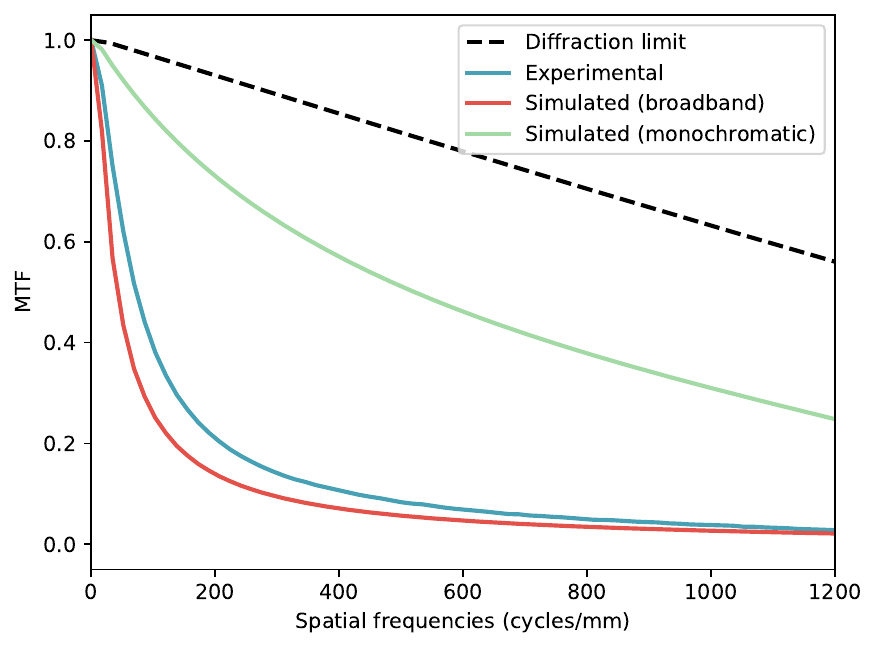}
    \caption{Experimental and simulated modulation transfer functions (MTF) of the hyperbolic metalens at 0$^\circ$ angle of incidence. The significant deviation from the diffraction limit is due to the broadband nature of the laser (FWHM $\approx$ \SI{5}{\nm}) used to measure the PSFs.}
    \label{fig:mtf}
\end{figure}

\newpage

\section{Imaging setup for hyperbolic metalens camera}
\begin{figure}[h]
    \centering
    \includegraphics[width=\textwidth]{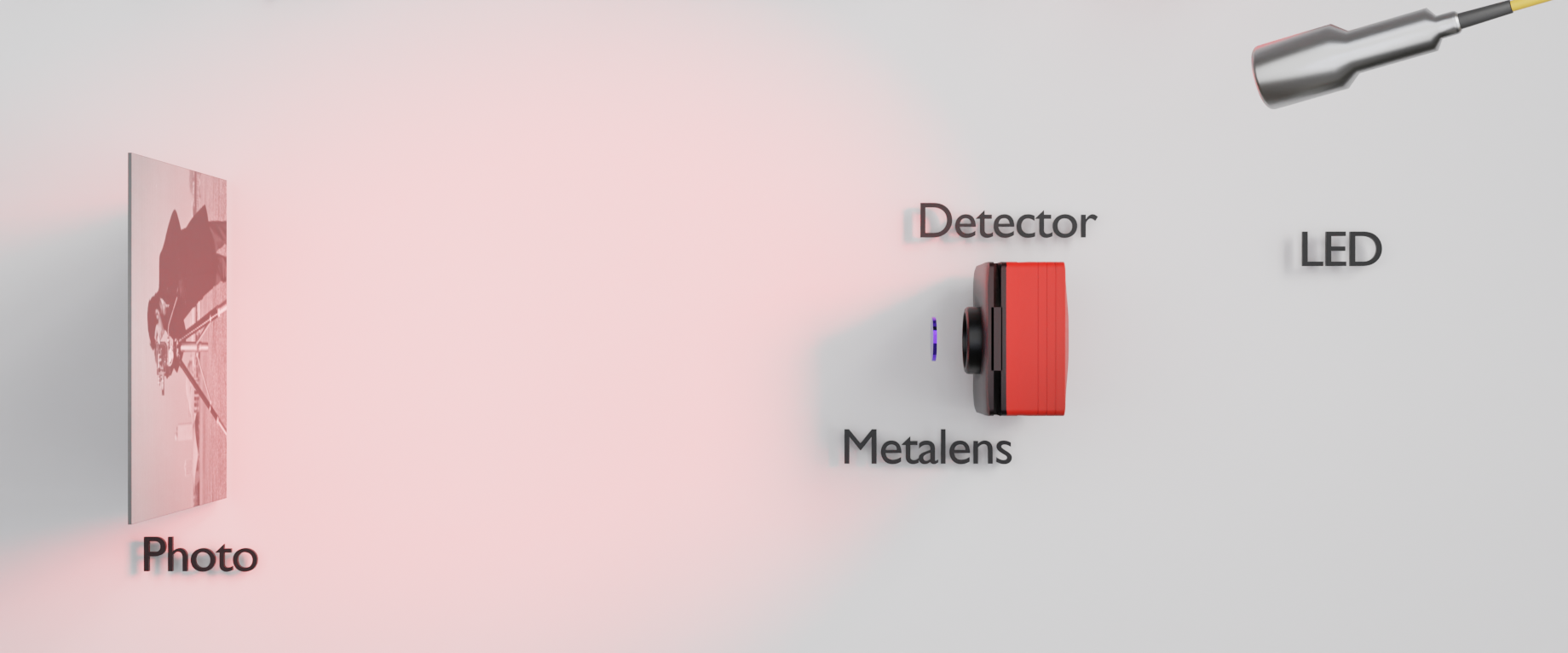}
    \caption{Imaging setup for the hyperbolic metalens camera. 
    The scene is illuminated with an LED with a dominant wavelength of \SI{850}{\nm} (Thorlabs M850L3) and bandwidth of \SI{30}{\nm}. The detector's outer casing is removed (not depicted here) to allow for direct mounting of the metalens a focal length away from the sensor (see Fig. 1g,h of the main manuscript).}
    \label{fig:imaging}
\end{figure}

\newpage

\section{Comparison of deblurring algorithms on simulated images}
\begin{figure}[!h]
    \centering
    \includegraphics[width=0.84\textwidth]{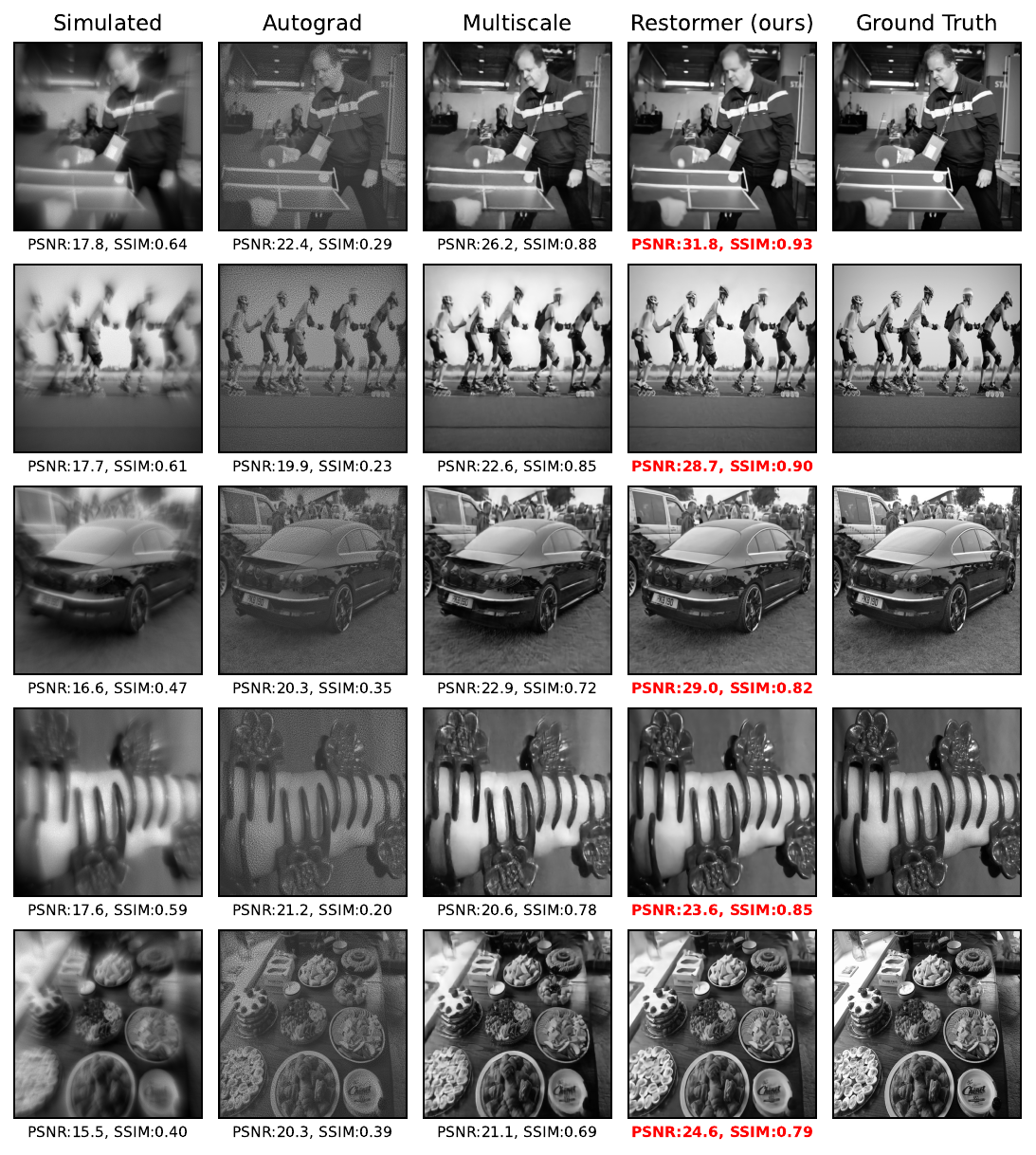}
    \caption{Comparison of deblurring algorithms on simulated images. The deblurred images' peak signal-to-noise ratio (PSNR) and structural similarity index measure (SSIM) are computed against the ground truth. The Restormer consistently achieves the highest PSNR and SSIM.}
    \label{fig:sim_compare}
\end{figure}

\newpage

\section{Restormer deblurring of experimental and simulated images}
\begin{figure}[h]
    \centering
    \includegraphics[width=\textwidth]{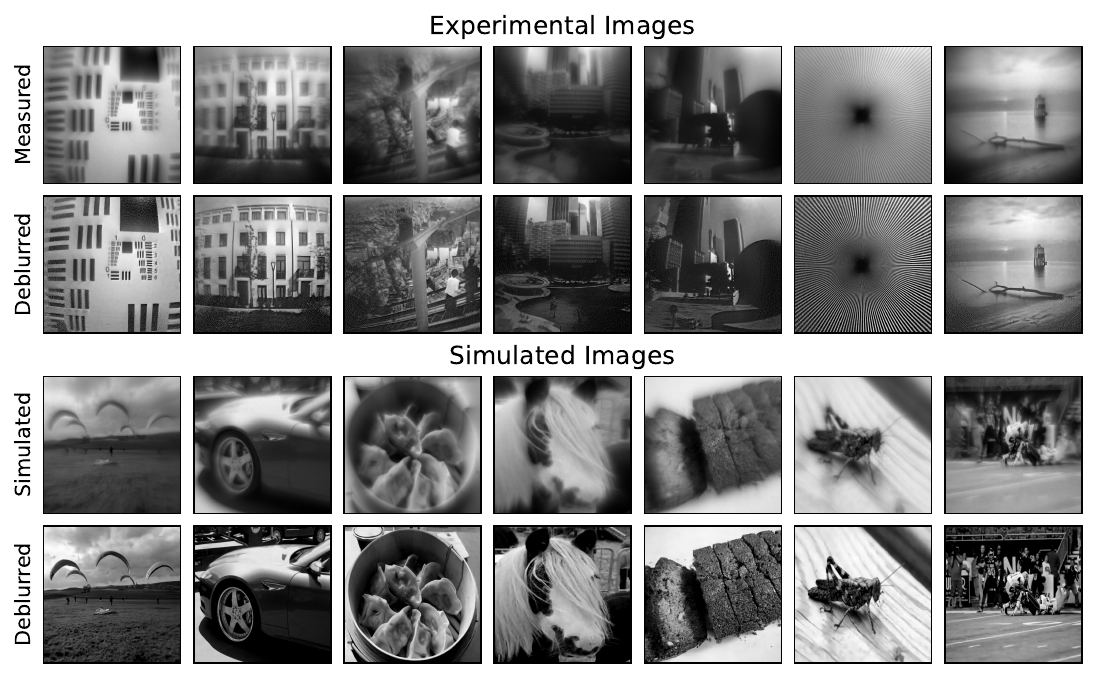}
    \caption{Restormer deblurring for both experimental and simulated images.}
    \label{fig:allimages}
\end{figure}

\section{Autograd implementation of the eigenCWD algorithm}
The eigenCWD algorithm \cite{Yeo2024-pm} seeks a solution that minimizes the following objective function based on the eigenPSF forward model described in Eq.~\eqref{eq:eigenconv}:
\begin{align}\label{eq:loss}
    \min_\mathbf{f} E\qty(\mathbf{f}) = \min_\mathbf{f} \frac{\mu}{2}\norm{\mathbf{g}_\text{eigenPSF}\qty(\mathbf{f}) - \mathbf{g}_\text{measured}}_2^2 + \alpha \norm{\nabla\mathbf{f}}_1,
\end{align}
where $E$ is the loss function, $\mathbf{f}$ and $\mathbf{g}$ denote the 2D-array representation of the object and blurred image respectively, $\mu$ and $\alpha$ are hyper-parameters, and $\nabla\mathbf{f}$ is the total variation regularizer applied on the object.
In the original implementation, the authors solve Eq.~\eqref{eq:loss} using the alternating direction method of multipliers (ADMM) method which required analytical gradients to be calculated.
This makes it tedious to change the loss function and regularizer as the analytical gradients have to be recalculated again.

Instead, we utilize PyTorch's automatic differentiation engine which uses the chain rule to computationally calculate the gradient update based on the differentiable, elementary functions used in the forward model of Eq.~\eqref{eq:eigenconv}.
This presents a highly flexible optimization method as one can easily change the loss metric or regularization function without the need to recalculate analytical gradients.

In the example shown in Fig. 5 of the paper, we used the following loss function:
\begin{align}
    \min_\mathbf{f} E\qty(\mathbf{f}) = \min_\mathbf{f} \mathrm{SSIM}\qty(\mathbf{g}_\text{eigenPSF}\qty(\mathbf{f}) ,\,\mathbf{g}_\text{measured}),
\end{align}
where the $\mathrm{SSIM}$ is the structural similarity index measure \cite{Wang2004-qe}.
We use PyTorch's inbuilt stochastic gradient descent (SGD) optimizer with default parameters and a learning rate of 100.
The code for this Autograd implementation of eigenCWD is available from the authors upon reasonable request.

\section{Multiscale Neural Network}
In this paper, we used the default parameters of the original Multiscale architecture \cite{Nah2017-ot}, except for changing the convolution kernel size to $7\times 7$, and the error metric from mean-squared error (MSE) to SSIM.
Through our empirical tests, these changes led to improvements in reconstructions.
We trained with a batch size of 64 for a total of 1000 epochs, which took approximately 7 hours to complete on 4 NVIDIA L40 GPUs.

\end{document}